\documentclass[%
reprint,
superscriptaddress,
 amsmath,amssymb,
prb,
]{revtex4-2}

\usepackage{graphicx}
\usepackage{dcolumn}
\usepackage{bm}

\begin{document}

\title{Damping-dependent thermalization of neighboring nanomechanical resonators below 1 mK}

\author{Amir Youssefi}
\thanks{These authors contributed equally to this work.}
\affiliation{Laboratory of Photonics and Quantum Measurement (LPQM), Swiss Federal Institute of Technology Lausanne (EPFL), Lausanne, Switzerland}
\affiliation{EDWATEC SA, EPFL Innovation Park Bâtiment D, 1015-Lausanne, Switzerland}
\author{Mahdi Chegnizadeh$^*$}
\affiliation{Laboratory of Photonics and Quantum Measurement (LPQM), Swiss Federal Institute of Technology Lausanne (EPFL), Lausanne, Switzerland}
\affiliation{Center for Quantum Science and Engineering, EPFL, Lausanne, Switzerland}
\affiliation{Institute of Physics, Swiss Federal Institute of Technology Lausanne (EPFL), Lausanne, Switzerland}
\affiliation{Institute of Electrical and Micro Engineering, Swiss Federal Institute of Technology Lausanne (EPFL), Lausanne, Switzerland}
\author{Francis Bettsworth}
\affiliation{Univ. Grenoble Alpes, CNRS, Grenoble INP, Institut Néel, 38000 Grenoble, France}
\author{Richard Pedurand}
\affiliation{Univ. Grenoble Alpes, CNRS, Grenoble INP, Institut Néel, 38000 Grenoble, France}
\author{Eddy Collin}
\affiliation{Univ. Grenoble Alpes, CNRS, Grenoble INP, Institut Néel, 38000 Grenoble, France}
\author{Tobias J. Kippenberg}
\affiliation{Laboratory of Photonics and Quantum Measurement (LPQM), Swiss Federal Institute of Technology Lausanne (EPFL), Lausanne, Switzerland}
\affiliation{Center for Quantum Science and Engineering, EPFL, Lausanne, Switzerland}
\affiliation{Institute of Physics, Swiss Federal Institute of Technology Lausanne (EPFL), Lausanne, Switzerland}
\affiliation{Institute of Electrical and Micro Engineering, Swiss Federal Institute of Technology Lausanne (EPFL), Lausanne, Switzerland}
\author{Andrew Fefferman}
\email{andrew.fefferman@neel.cnrs.fr}
\affiliation{Univ. Grenoble Alpes, CNRS, Grenoble INP, Institut Néel, 38000 Grenoble, France}

\date{\today}

\begin{abstract}
The position noise spectra of six drums on a single chip were measured on a single cooldown below 1.3 kelvin. Cryostat temperatures as low as 0.7 mK were achieved. The temperature dependence of the resonance frequency and linewidth of the drum modes was analyzed in the framework of the tunneling two level system (TLS) model. Departures of the resonance frequency and the position noise power from the expected logarithmic and linear temperature dependences, respectively, were interpreted as indications of thermal decoupling from the cryostat. This previously unexplored measurement configuration revealed that similar neighboring drums on a single chip may be at different temperatures. At the lowest temperatures, some drums exhibited excess damping that decreased with temperature. The magnitude of the excess damping of the drums was correlated with the thermal coupling of their TLS to the cryostat. In the case of one drum, a temporary increase in its damping coincided with a decrease in its mode temperature. The thermalization of the TLS to the cold finger was independent of pump power, pulse tube state and temperature of the pre-cooling stages of the cryostat. These results suggest that thermalization of nanomechanics to the cryostat can be mediated by TLS damping rather than clamping loss and may impact efforts to extend the coherence of mechanical resonators.
\end{abstract}

\maketitle

\section{\label{sec:intro} Introduction}
Understanding environmental influences on quantum systems is a ubiquitous problem. In some cases, coupling to external degrees of freedom is desirable for technological applications or fundamental interest. This includes quantum sensing \cite{degen2017quantum} and thermodynamics experiments \cite{binder2018thermodynamics} including quantum engines \cite{scovil1959three,scully2003extracting}. In other cases, one wishes to minimize coupling to noise sources so as to improve the system's coherence, such as in quantum computation.

The effective temperature of superconducting electronic devices typically saturates at 30-50 mK when cooled by dilution refrigerators reaching 10 mK. This has been modeled in terms of coupling of the device to multiple baths at different temperatures \cite{lvov2025thermometry}. The baths may be global, affecting all devices in the cryostat, or they may locally influence individual devices \cite{sharafiev2025leveraging}. At the lowest temperatures, thermalization of electronic systems to the cryostat is limited by weak electron-phonon coupling \cite{wang2019crossover}.

Special measures can be taken to improve the thermalization of the electrons to the cryostat below the $\approx30$ mK limit. These include increasing the volume in which electron-phonon coupling takes place \cite{bradley2016nanoelectronic} as well as nuclear demagnetization of the device wiring \cite{sarsby2020500}. On-chip nuclear demagnetization is another technique for cooling electronic devices that is suitable for samples the can tolerate maximum magnetic fields of several tesla \cite{samani2022microkelvin}. Immersion in superfluid $^3$He, in contrast, is better adapted to typical superconducting circuits. When applied to superconducting resonators, this technique has been shown to decrease the saturation temperature of $1/f$ noise from 80 mK to the microkelvin range and to reverse the usual increase of noise magnitude with cooling \cite{lucas2023quantum}.

Nanomechanical systems are also potentially applicable to quantum technologies \cite{pistolesi2021proposal}. Due to their small size and long coherence times \cite{pechal2018superconducting}, they can serve as microwave quantum memories working in conjunction with electronic systems \cite{bozkurt2025mechanical}. Concerning fundamental physics, relatively massive mechanical resonators can have decoherence times comparable to the predicted gravitational decoherence time, opening the possibility of relativistic tests of quantum mechanics \cite{gerashchenko2025probing}. Furthermore, detection of single gravitons with highly coherent MHz frequency mechanical resonators has been proposed \cite{tobar2024detecting}, potentially progressing beyond detection of gravitational waves consisting of very large numbers of gravitons  \cite{abbott2016observation}. Estimated coherence times exceeding 100 msec have been observed in electromechanical systems at 30 mK \cite{seis2022ground}. The continual decrease in the mechanical damping observed in \cite{seis2022ground,youssefi2023squeezed} as the dilution refrigerator approached its base temperature suggests that it could be possible to achieve even longer coherence times by cooling to sub-mK temperatures. Such refrigeration could also enhance the mechanical coherence time by decreasing the equilibrium phonon occupation of the mechanical mode. Pioneering experiments with 10 MHz aluminum drum modes demonstrated a minimum occupancy of 30 mechanical quanta while in thermal equilibrium with the cryostat at 15 mK \cite{teufel2011sideband}.

Thermalizing nanomechanics to microkelvin temperatures has been demonstrated using conventional off-chip adiabatic nuclear demagnetization with neither on-chip nuclear demagnetization nor immersion in superfluid $^3$He \cite{cattiaux2021macroscopic}. In contrast to the electronic case, the mechanical system benefits from weak electron-phonon coupling: it serves to insulate the mechanical system from hot electrons in the device wiring. The phonons of the sample can then closely approach the temperature of the phonons of the cryostat's lowest temperature stage.

Thermalization of nanomechanics below 1 mK is of interest not only for achieving long coherence times but also for studying TLS. If cooling of the mechanics results in growth of the dominant thermal phonon wavelength beyond a sample dimension, changes in the effects of TLS on the mechanical properties are expected \cite{behunin2016dimensional}. Furthermore, cooling the mechanical mode near the ground state allows detection of individual TLS, and sub-mK temperatures may be required in the case of MHz frequency mechanics \cite{remus20009damping,pedurand2024progress,yuksel2025intrinsic}. Cooling nanomechanics to ultra-low temperatures also enables studies of phononic thermodynamics \cite{golokolenov2023thermodynamics,van2025sub}.

We measured the temperature dependence of the resonance frequency and the damping of several similar drums on a single chip. The measurements were made on a single run where the cryostat's base plate temperature reached 0.7 mK and did not exceed 1.3 kelvin. Measuring numerous neighboring nanomechanical resonators at these ultra-low temperatures was previously unexplored. It revealed variations in the behavior of the mechanical modes, whose resonances ranged from 1.3 to 2.4 MHz.  In this work, we focus on six devices whose resonance frequency decreased logarithmically with temperature upon cooling from 200 mK to a device-dependent saturation temperature. The saturation temperature varied by a factor of 10 among the different drums. This variation indicates that the frequency shift of each drum is caused by an ensemble of TLS with a unique minimum temperature. This minimum temperature is determined by the coupling of each TLS ensemble to the cryostat and other, warmer baths. We rule out several candidates for the dominant warm bath, including some that limit the cooling of superconducting qubits. The minimum temperature of a mechanical mode was inversely correlated with its temperature-dependent damping. This suggests that TLS damping, rather than clamping losses, were responsible for thermal coupling of the mechanical mode to the cryostat.

\section{Theory}
\label{sec:theory}
The effects of TLS on nanomechanics generally depend on the dispersion relation of the mode and the phonon density of states. However, the dissipation due to resonant absorption of phonons by TLS in the limit of low oscillating strain remains proportional to the bulk expression \cite{behunin2016dimensional}. In particular, the damping rate is
\begin{equation}
\label{eq:damp}
\Gamma_{\mathrm{res}}=\pi\omega C\tanh\left(\frac{\hbar\omega}{2k_BT}\right)
\end{equation}
where $C$ is the tunneling strength, $\omega$ is the angular frequency of the phonons and $T$ is the temperature. In the case of a bulk sample, the tunneling strength $C_\mathrm{bulk}=P_0 \gamma^2/Y$, which is a ratio of microscopic TLS parameters to the elastic modulus, $Y$, of the material. In particular, $P_0$ is the TLS spectral density prefactor and $\gamma$ is the deformation potential corresponding to the coupling between TLS and phonons. In the case of a string, $C_{\mathrm{string}}=P_0\gamma^2\pi^2e^2/12\sigma_0l^2$ for thickness of the string $e$, built-in axial stress $\sigma_0$ and string length $l$ \cite{maillet2023nanomechanical}. We expect the tunneling strength in a drum to depend similarly on its aspect ratio and built-in stress. Stress causes dissipation dilution, but we do not expect stress to affect the distribution of TLS parameters or the coupling of TLS to phonons \cite{ftouni2015thermal}. The variation of the sound speed due to resonant interactions between TLS and phonons is obtained from the resonant damping rate by the Kramers-Kronig relation \cite{phillips1987two}, yielding a change in resonance frequency $f_m-f_0$ relative to a reference frequency $f_0$ at temperature $T_0$
\begin{equation}
\label{eq:df}
\left.\frac{f_m-f_0}{f_0}\right|_{res}=C\ln\left(\frac{T}{T_0}\right)
\end{equation}
for $k_BT\gg\hbar\omega$.

In previous work, the damping of phonon pulses due to resonant TLS predicted by Eq. \ref{eq:damp} was observed near 1 GHz and 20 mK \cite{golding1976intrinsic}. The samples were non-resonant 6 mm cubes of fused silica, and the acoustic intensities were kept low enough so that the TLS were not saturated by strain. More recently, this effect was observed via Brillouin-Mandelstam scattering in optical fibers down to 50 mK \cite{cryer2025brillouin}.

In nanomechanical devices, the zero-point strain can exceed the strain $\epsilon_{\mathrm{sat}}=\hbar/\gamma\sqrt{\tau_1\tau_2}$ expected to saturate resonant TLS, where $\tau_1$ is the TLS relaxation time, $\tau_\phi$ is its dephasing time, and $\tau_2^{-1}=(2\tau_1)^{-1}+\tau_{\phi}^{-1}$ \cite{remus20009damping}. From parameters obtained from measurements of Al films at kHz frequencies \cite{fefferman2010elastic}, a TLS resonant with our mechanical modes near 1.5 MHz is expected to have $\tau_1>20$ sec at 10 mK and $\gamma=3.8$ eV \cite{pedurand2024progress}. Assuming $\tau_2\approx10~\mu$sec \cite{remus20009damping}, we obtain $\epsilon_{\mathrm{sat}}\approx10^{-14}$. The parameters $\tau_1$ and $\tau_2$ could be substantially modified by the small length scales of our drums (see below). We can compare $\epsilon_{\mathrm{sat}}$ with the zero point strain of the drums of the present work. In analogy with the strain due to flexure of a beam \cite{cleland2003foundations}, the zero point strain at the center of the surface of a drum is approximately $2 x_{\mathrm{zp}}h\partial^2\psi/\partial r^2$. Here $x_{\mathrm{zp}}=1$ fm is the approximate zero-point motion of our drums, $h=100$ nm is half the thickness of a drum and $\psi$ is the shape of the fundamental mode \cite{youssefi2023squeezed}. The curvature of the fundamental mode shape at its center is $0.5 (\lambda_{0,0}/R)^2$, where $\lambda_{0,0}=2.40$ and the radius of the drum $R\approx100 \mu$m \cite{cattiaux2020geometrical}. Thus the zero-point strain due to bending is of order $6\times10^{-14}$. The zero-point strain due to elongation of the drum is of order $(x_{\mathrm{zp}}/R)^2=10^{-22}$. We conclude that resonant TLS in our drums may be saturated even for the lowest phonon occupation of the mechanical modes.

Damping due to relaxation of TLS is typically more important than resonant damping when $k_BT\gg\hbar\omega$ is satisfied, as in the present work. This contribution to damping is not saturable and is substantially modified by reduced dimensionality and phonon dispersion \cite{behunin2016dimensional}. In particular, phonon-driven relaxation of TLS is expected to produce damping with a cubic temperature dependence in bulk insulating materials in the low temperature limit \cite{phillips1987two}, while in superconducting or insulating one-dimensional strings \cite{maillet2023nanomechanical}, one-dimensional half rings \cite{hauer2018two}, and one- and two-dimensional cantilevers \cite{kamppinen2022dimensional}, departures from the $T^3$ dependence are expected and observed. In the case of 1 to 4 MHz modes of insulating or superconducting membranes, the damping between 20 mK and 500 mK is typically well described by a linear temperature dependence plus a constant offset \cite{brubaker2022optomechanical,suh2013optomechanical,Wollman_thesis}. A notable exception is the membrane studied in \cite{seis2022ground}, whose damping had a $T^{0.63}$ dependence with no discernible offset down to 30 mK. The theoretical damping due to TLS relaxation and $\epsilon_{\mathrm{sat}}$ both depend on $\tau_1^{-1}$. Deriving this relaxation rate due to phonons in membranes of reduced dimensionality is beyond the scope of the present work.

\section{Experiment}
\label{sec:expt}
The sample consisted of several aluminum, lumped element microwave resonators (cavities) with a mechanically compliant capacitor plate (drum), all arranged along a central transmission line. The procedure for fabricating samples of this type is described in \cite{youssefi2025compact}. The diameters of the drums on the chip ranged from 110 to 180 micrometers and their film thickness was 200 nm. The typical average tensile stress due to deposition conditions and differential thermal contraction is 400 MPa in the low temperature limit \cite{youssefi2025compact}. The local stress in the clamps of a drum increases with the ratio of the etch perimeter to the sum of the widths of the twelve clamps \cite{bereyhi2019clamp,youssefi2025compact}. This clamp ratio had an average value of 2 and varied by no more than a factor of 2.4 between drums.

\begin{figure}
\includegraphics[width=\linewidth]{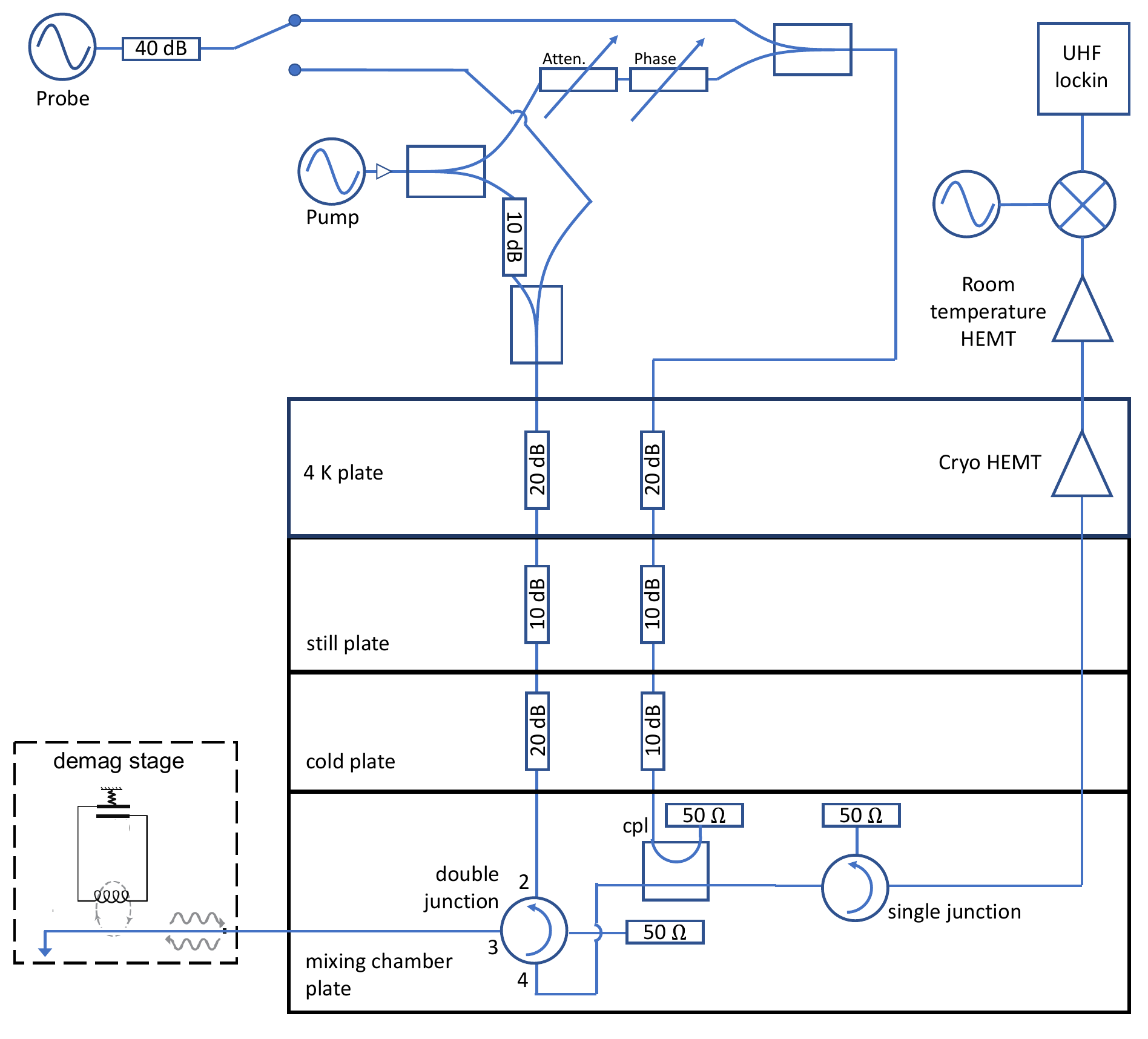}
\caption{\label{fig:circuit} A Keysight N5173B generator was used to pump the optomechanical device. One channel of an Anapico APMS12G generator was used as a microwave probe and the other channel was used to mix the signal reflected by the sample down to the low MHz range. A Zurich UHF lockin amplifier was then used to measure the noise power spectrum centered on the upper mechanical sideband of the reflected signal. Inside the cryostat, the indicated attenuators along with the 10 dB loss of the directional coupler yielded 50 dB of attenuation distributed between the 4 kelvin plate and the 100 mK cold plate. This attenuation and the isolation provided by the circulators protected the sample from noise. Initial amplification of the output signal was provided by a cryogenic Low Noise Factory LNC4\_8C. Further amplification at room temperature was provided by a Low Noise Factory LNR4\_8C.}
\end{figure}

Devices of this type were previously shown to have damping rates as low as 40 mHz at dilution refrigerator base temperature, yielding a phonon lifetime $T_1=7.7$ msec, with coherence nearly unaffected by pure dephasing \cite{youssefi2023squeezed}. The high-yield fabrication of these devices simultaneously achieved long mechanical coherence times and relatively high optomechanical coupling, as well as the capacity for strong microwave pumping without excessive non-equilibrium population of the cavity \cite{youssefi2025compact}. The motion of the drums, which was largely due to thermal agitation, was detected with microwave optomechanics techniques \cite{aspelmeyer2014cavity}. We used a standard microwave circuit for this purpose (Fig. \ref{fig:circuit}). The microwave pump frequency was aligned with the resonance of a microwave cavity, so that the motion of its drum produced sidebands in the reflected signal. Thermal motion of the drum produces Lorentzian sidebands separated from the pump frequency by $\pm f_m$ and having the linewidth $\Delta f$ of the mechanics. The pump power was low enough so that dynamical backaction did not affect $\Delta f$ or $f_m$.

A unique aspect of the experimental setup was the use of cryogen-free nuclear demagnetization refrigeration to reach sub-mK cryostat temperatures. This technique relies on an additional stage of refrigeration that is pre-cooled by a standard cryogen-free dilution refrigerator. Microwave optomechanics cooled by conventional (``wet") nuclear demagnetization was first reported in \cite{zhou2019chip}. In the present work, a copper sample holder of the type described in \cite{youssefi2023squeezed,youssefi2025compact} was pressed against a cold finger extending from the refrigerator described in \cite{raba2024aluminum}. On a previous cooldown of the same type of sample holder, a noise thermometer \cite{MFFT,engert2012noise} was attached to the copper plate used to press the sample holder to the cold finger, yielding an upper limit of 1.5 mK for the temperature of the sample holder at the base temperature of the cryostat.

\begin{figure}
\includegraphics[width=\linewidth]{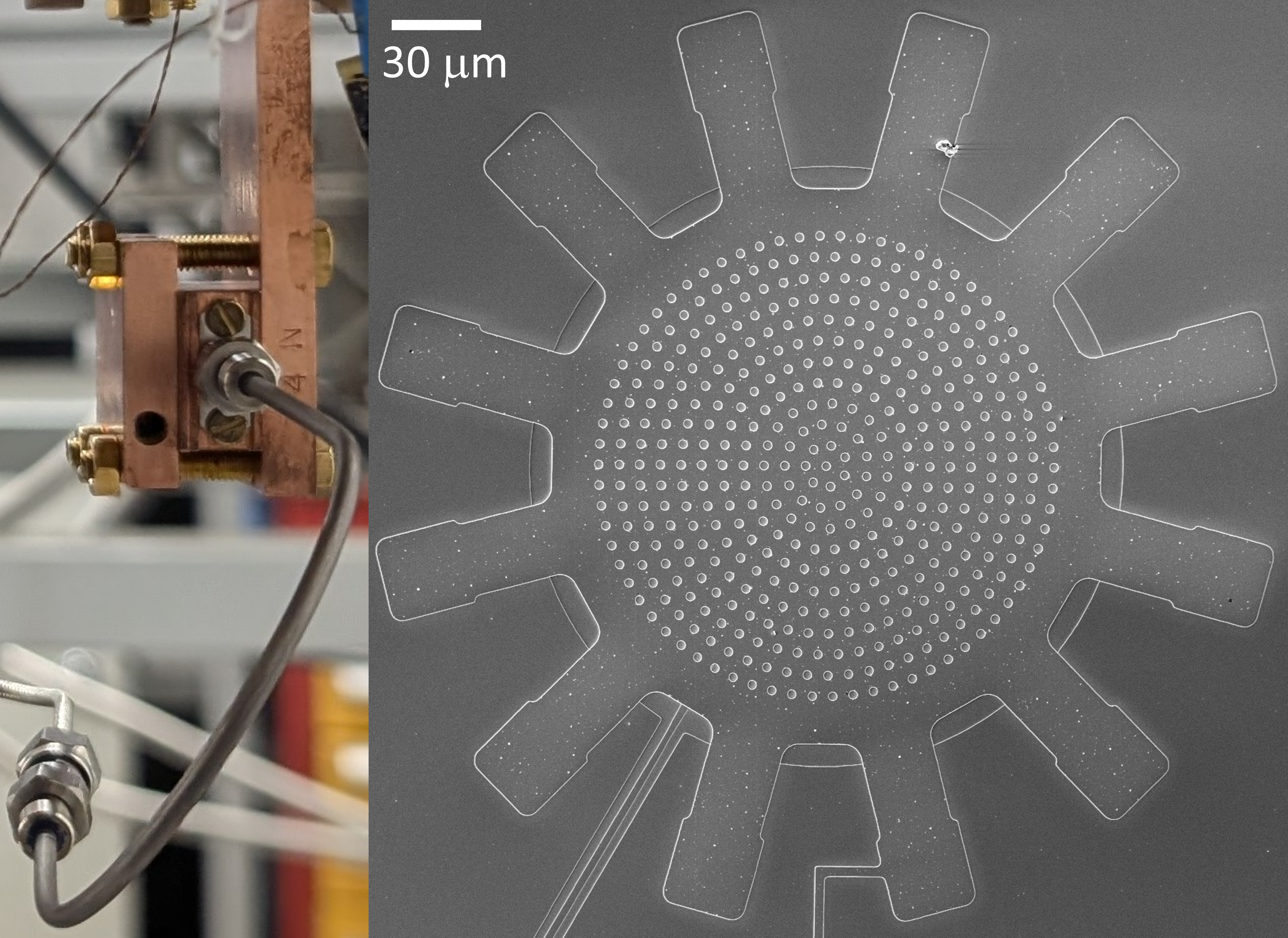}
\caption{\label{fig:cell-photo} Left panel: The sample holder clamped to the cold finger of the nuclear stage. The NbTi coaxial line transmits microwave signals to and from one port of the sample holder. An SMA connector attached to the other port (not visible) terminates the microwave circuit with a short to ground. Right panel: An SEM image of one of the drums on the chip. It is suspended above a fixed electrode (not visible) that was deposited in a hole in the substrate. The circular edge of the hole is visible between each of the twelve clamps of the drum. The small holes in the drum facilitate penetration of the etchant in the release step of fabrication  \cite{youssefi2025compact}.}
\end{figure}

Special measures were taken to prevent parasitic heating of the sample. In particular, a 15 cm length of superconducting NbTi coaxial line allows low-loss transmission of microwaves to and from the sample while maintaining its thermal isolation from the mixing chamber of the dilution refrigerator (Fig. \ref{fig:cell-photo}). Furthermore, millimeter thick copper plates with a total surface area of order 1000 cm$^2$ were covered in black paint or paper tape and attached to the mixing chamber plate so as to absorb stray infrared photons. These photons, which are reflected by shiny metallic surfaces of the cryostat, could have leaked inside the standard 1 kelvin radiation shield through holes in the still plate that pass wires.

\section{Results}
\subsection{Thermalization of TLS}
\label{sec:ThermTLS}
\begin{figure}
\includegraphics[width=\linewidth]{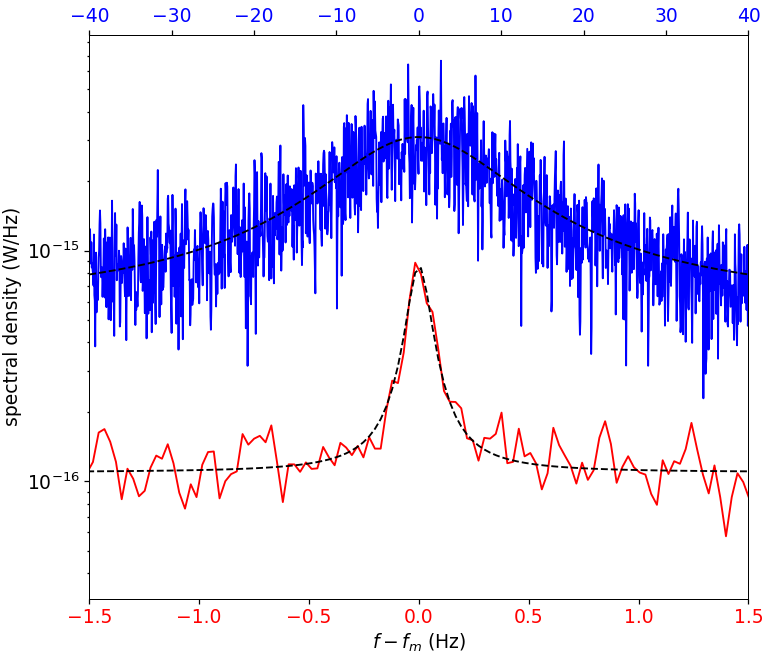}
\caption{\label{fig:spectra} Noise power spectral density due to vibrational modes of two different drums at 1.87 MHz (upper trace and upper axis, blue) and 1.68 MHz (lower trace and lower axis, red). The cryostat temperatures were 0.8 and 0.7 mK, respectively. The background level is much lower in the lower trace because a filter was inserted at room temperature on the output port of the cryostat. Dashed black curves are Lorentzian fits.}
\end{figure}

\begin{figure}
\includegraphics[width=\linewidth]{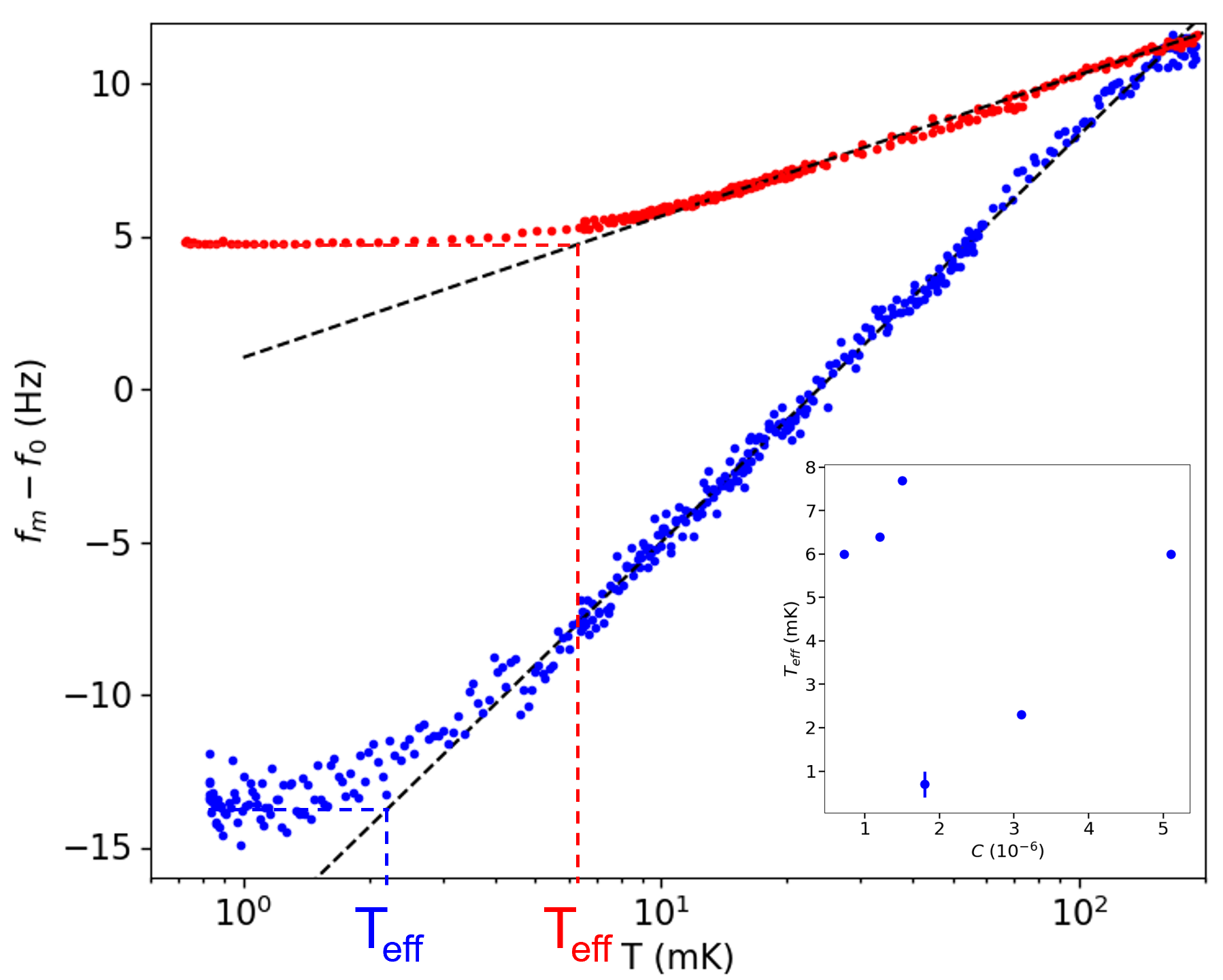}
\caption{\label{fig:f-vs-T} Resonance frequency shift of 1.87 MHz (blue) and 1.68 MHz (red) drums obtained from spectra like those shown in Fig. \ref{fig:spectra}. Black dashed lines correspond to Eq. \ref{eq:df}. Dashed red and blue lines indicate the method for determining the saturation temperature $T_{\mathrm{eff}}$. Inset: No systematic dependence of $T_{\mathrm{eff}}$ on $C$.}
\end{figure}

\begin{figure}
\includegraphics[width=\linewidth]{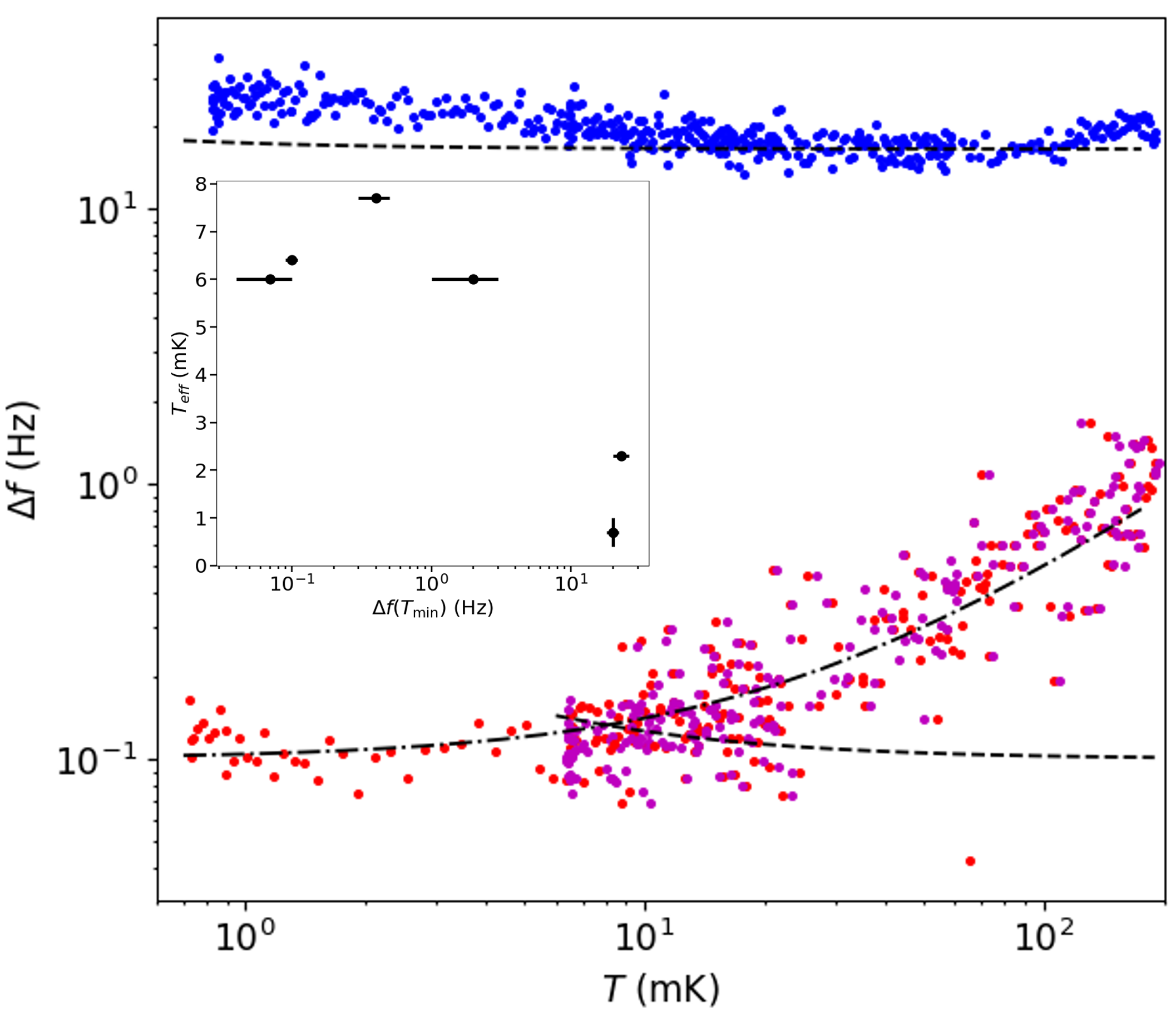}
\caption{\label{fig:damping-vs-T} Mechanical linewidth of 1.87 MHz (blue) and 1.68 MHz (red) drums obtained from spectra like those shown in Fig. \ref{fig:spectra}. The magenta points were obtained from the red ones by accounting for thermal decoupling of the sample from the thermometer (Sec. \ref{sec:ThermTLS}). Dashed curves correspond to the resonant contribution to the damping (Eq. \ref{eq:damp}) based on $C$ obtained from the resonance frequency shift plus a constant offset. The dash-dot curve is a linear fit with slope 4.04 Hz/K and offset 0.10 Hz. Inset: Dependence of $T_{\mathrm{eff}}$ on mechanical linewidth at base temperature for six different drums.}
\end{figure}

Figure \ref{fig:spectra} shows the upper mechanical sideband from two different drums near the base temperature of the cryostat. The figure illustrates the large variation in drum damping we observed. A Lorentzian function plus a constant offset was fitted to the spectra to determine the linewidth $\Delta f$ and resonance frequency $f_m$. Figure \ref{fig:f-vs-T} shows the cryostat temperature dependence of $f_m$ and Fig. \ref{fig:damping-vs-T} shows the corresponding temperature dependence of $\Delta f$ of the two drums.

We fit the theoretical $f_m-f_0$ (Eq. \ref{eq:df}) to the high temperature part of the data shown in Fig. \ref{fig:f-vs-T}, obtaining values for the tunneling strength $C$. We characterized the leveling off of the frequency shift at the lowest temperatures by a temperature $T_{\mathrm{eff}}$, which corresponds to the temperature at which the value of the fit function equals the measured $f_m-f_0$ at the base temperature of the cryostat.

We measured four other drums whose frequency shift had a temperature dependence similar to the one shown in Fig. \ref{fig:f-vs-T}, i.e., a $\log T$ dependence at relatively high temperatures and a leveling off at the lowest temperatures. We followed the same procedure to extract $C$ and $T_{\mathrm{eff}}$. The tunneling strength $C$ was as low as $7\times10^{-7}$, which is well below the values near $2\times10^{-5}$ observed in the string studied in \cite{maillet2023nanomechanical} and in similar structures referenced in that work. The $T_{\mathrm{eff}}$ of all six drums is plotted against $C$ in the inset of Fig. \ref{fig:f-vs-T} and against the linewidth at 800 $\mu$K, $\Delta f(T_{\mathrm{min}})$, in Fig. \ref{fig:damping-vs-T}. While $T_{\mathrm{eff}}$ has no systematic dependence on $C$, it decreases with $\Delta f(T_{\mathrm{min}})$. 

The observed low temperature saturation of $f_m-f_0$ could be the intrinsic temperature dependence of the mechanical modes or it could be caused by thermal decoupling of the TLS from the cryostat. In some previous ultra-low temperature measurements of glass, the former possibility was supported (as explained below), but we believe the situation is different in the present work. The predictions of the tunneling model have the same form for the resonant contribution to the change in sound speed and dielectric constant \cite{behunin2016dimensional}. Rogge \emph{et al.} measured the dielectric constant of five types of glass down to 0.5 mK \cite{rogge1997nonlinear}. In order to ensure good thermalization, the samples were immersed in $^3$He and the electrodes for the capacitance measurements were thermalized to the $^3$He using silver powder heat exchangers. Despite the similarity of the environments of the samples, the dielectric constants of the samples saturated at different temperatures ranging from 8 mK to less than 1 mK, independent of the drive level. The variation in the saturation temperatures was thus taken to be intrinsic behavior and explained in terms of a material-dependent low energy cutoff of the matrix element $\Delta_0$ for transitions between configurations of the TLS. The saturation temperature increased monotonically with the magnitude of the slope of the higher temperature $\ln T$ dependence. However, we do not observe such a correlation (Fig. \ref{fig:f-vs-T}). Furthermore, mechanical measurements of the material that had the highest dielectric saturation temperature (silicon oxide, 8 mK) showed no sound speed saturation down to 2 mK without immersion in $^3$He \cite{fefferman2008acoustic}.

We therefore believe the considerations in \cite{rogge1997nonlinear} are not directly applicable to the present mechanical measurements, and that the variable saturation temperatures we observe can be explained by variation in the thermal coupling of the TLS to the cryostat. Under this assumption, we accounted for the effects of thermal decoupling on the temperature dependence of $\Delta f$ for the drum in Fig. \ref{fig:damping-vs-T} that was more strongly decoupled (1.68 MHz). In particular, we used the fit of Eq. \ref{eq:df} to the high temperature part of that drum's frequency shift (Fig. \ref{fig:f-vs-T}) to assign a corrected temperature to each data point. The result is shown by the magenta points in Fig. \ref{fig:damping-vs-T}.

One possible explanation for the dispersion in the $T_{\mathrm{eff}}$ across the drums is variation in the heat release from TLS in the aluminum \cite{nittke1996low}. Since the drums have approximately the same size, an increase in heat release could be due to an increase in the TLS density $P_0$, and consequently an increase in $C$ if $\gamma$ remains constant. However, the lack of a dependence of $T_{\mathrm{eff}}$ on $C$ suggests that if $P_0$ varies, it does not have a clear effect on thermal decoupling due to TLS heat release. On the other hand, the low $T_{\mathrm{eff}}$ observed for high $\Delta f(T_{\mathrm{min}})$ could be due to enhanced relaxation of a subset of TLS (Section \ref{sec:TLSdamping}). This subset of TLS could then help thermalize the TLS responsible for the mechanical frequency shift to the cryostat. TLS are known to interact with each other via strain, resulting in TLS spectral diffusion \cite{phillips1987two} and relaxation \cite{burin1995low}.

If the thermal transport between the substrate and the TLS probed by measuring $f_m$ is mediated by phonons rather than TLS interactions, then thermal decoupling of the drum modes from the substrate could have a significant effect on the TLS temperature. The low temperature thermal resistance between metal films and their insulating substrate contains a boundary contribution that has been modeled by mismatch theory. Following Refs. \cite{wang2019crossover,autti2023thermal}, we estimate a boundary resistance $R_{\mathrm{Al-Si}}AT^3=10^{-3}$ K$^4$m$^2$W$^{-1}$ between our Al drum and the Si substrate. For a contact area of order 10$^{-8}$ m$^2$ and a heat release of 20 nW/kg after 300 h in the low mK range \cite{nittke1996low}, we expect a negligible offset of 30 $\mu$K at 1 mK average temperature.

The clamping loss could provide an alternate route to determining the thermal resistance between the phonons of a drum and the substrate. The exponential suppression of the clamping loss in the azimuthal harmonic index \cite{wilson2011high} suggests that the fundamental mode is primarily responsible for thermal coupling of the drum to the substrate via the clamps. Furthermore, the modes of the drum may not be in thermal equilibrium due to weak phonon-phonon interactions \cite{polanco2023phonon}. In the present work, the upturn in the damping sometimes observed at the lowest temperatures prevents us from determining the clamping loss.

\subsection{TLS damping} \label{sec:TLSdamping}
Having obtained $C$ from the resonance frequency shift of a drum, we can consider the expected contribution of resonant damping to $\Delta f$. We noted that the thermal motion of our drum could be sufficient to saturate the resonant damping (Sec. \ref{sec:theory}). It is nonetheless interesting to compare the unsaturated resonant damping to our measurements. The dashed curves in Fig. \ref{fig:damping-vs-T} correspond to $\Gamma_{\mathrm{res}}/(2\pi)$ given by Eq. \ref{eq:damp} plus a constant offset that could be due to clamping loss. The unsaturated resonant damping is not large enough to account for the upturn in the linewidth observed at the lowest temperatures in the 1.87 MHz mode. The dashed curve for the 1.68 MHz mode only extends down to 6 mK since we believe that the TLS do not cool below that temperature on this drum. Resonant damping does not seem to play an important role in these devices.

In the present work, the damping of some drums is well described by a linear temperature dependence plus an offset (Figs. \ref{fig:damping-vs-T} and \ref{fig:PT-damping}). That is consistent with measurements of membranes in several other works (Sec. \ref{sec:theory}). However, in other drums of the present work, the damping at temperatures below 10 mK increases with cooling (Fig. \ref{fig:damping-vs-T}). The latter case is incompatible with the phonon-driven relaxation damping that occurs in glasses with a broad distribution of TLS energies, since the resulting damping decreases monotonically with cooling. However, a system of defects with a particular energy splitting yields a dissipation peak at the temperature where the relaxation rate of the defects matches the angular drive frequency (Debye relaxation). It also yields a decrease in stiffness as the sample warms through the dissipation peak and the defects begin to contribute to the mechanical response. A surplus of TLS at a certain energy could therefore produce the observed upturn in the damping and contribute to the leveling off of the resonance frequency on cooling. The detailed origin of this upturn, the pronounced temporary upturn in the damping in another drum (Fig. \ref{fig:area-vs-T}), and damping peaks observed in other drums on the same chip will all be the subject of a future article.

\subsection{Potential heat sources}
\begin{figure}
\includegraphics[width=\linewidth]{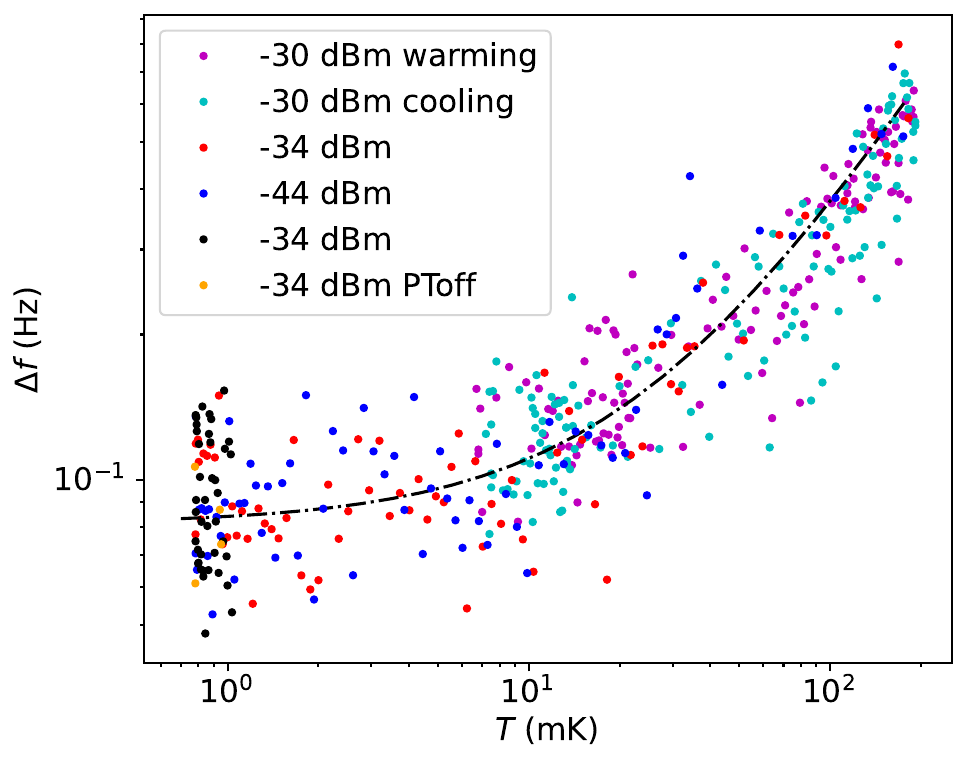}
\caption{\label{fig:PT-damping} Linewidth of a drum vibrating at 1.30 MHz under various measurement conditions. The settings of the pump generator are indicated in the legend. In order to maintain the stray magnetic field of the nuclear demagnetization refrigerator at the sample at a constant level below 100 $\mu$T, the drum measurements reaching 0.8 mK (red, blue, black and orange points) were made on warming. The orange data points were obtained below 1 mK with the pulse tube turned off. The dash-dot curve is a linear fit with slope 2.97 Hz/K and offset 0.08 Hz.}
\end{figure}

\begin{figure}
\includegraphics[width=\linewidth]{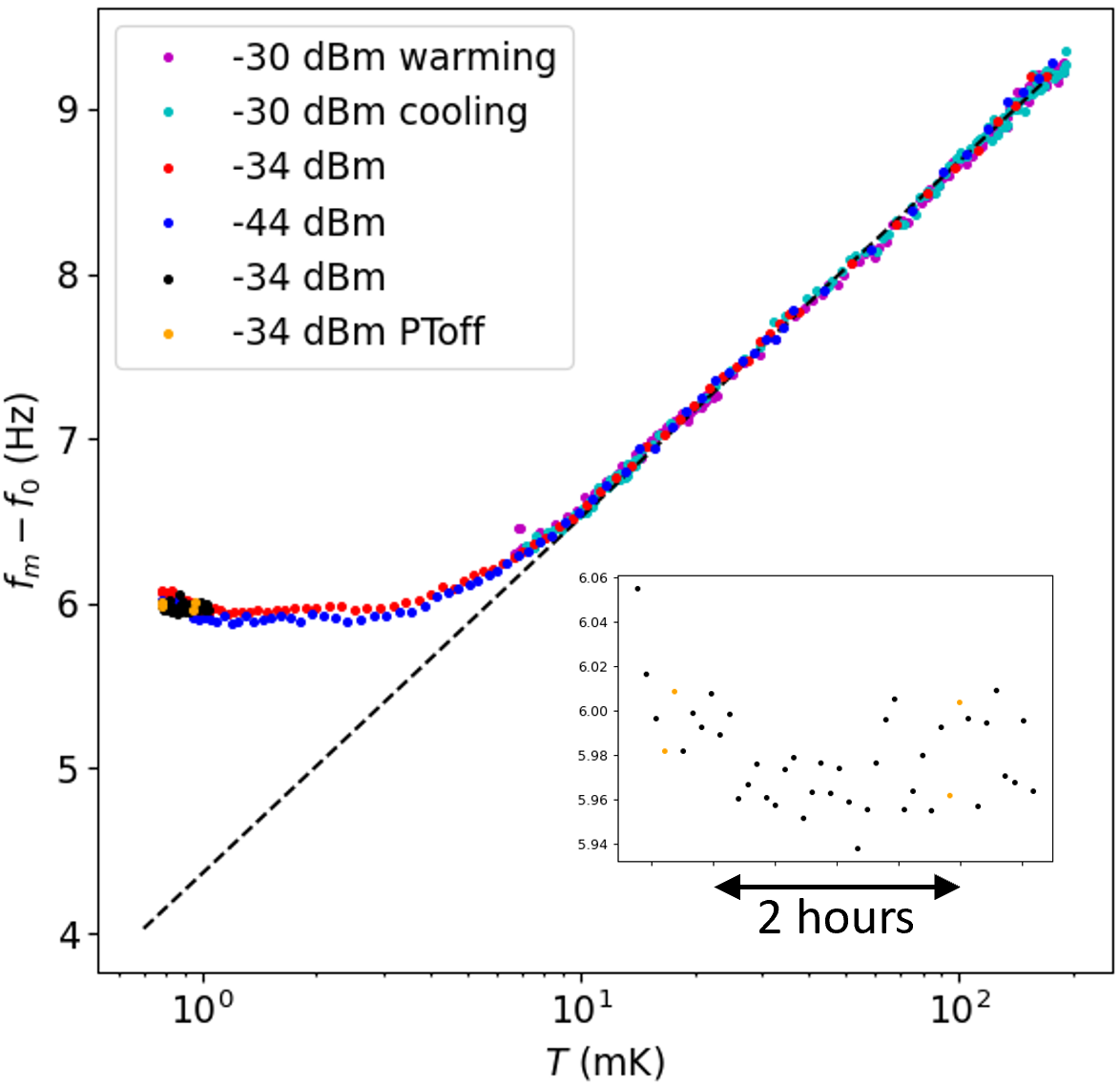}
\caption{\label{fig:PT-f} Frequency shift of a drum vibrating at 1.30 MHz under various measurement conditions. See the Fig. \ref{fig:PT-damping} caption for an explanation of the legend entries. The black dashed line corresponds to Eq. \ref{eq:df}. Inset: time dependence of the frequency shift highlighting the lack of a dependence on the pulse tube state below 1 mK.}
\end{figure}

We now demonstrate the reproducibility of the measurements and show that several potential heating mechanisms are not significant in our case. Figures \ref{fig:PT-damping} and \ref{fig:PT-f} respectively show the temperature dependence of $\Delta f$ and $f_m$ of another drum vibrating at 1.30 MHz. The figures show measurements on both warming and cooling down to the base temperature of the dilution refrigerator near 6 mK, demonstrating the absence of hysteresis. We observed occasional jumps in $f_m$ that on average occurred less than once per temperature sweep. We added offsets to $f_m$ so that the jumps are not visible in the plots. Aside from these jumps and a temporary transition (Fig. \ref{fig:area-vs-T}), the $f_m$ and $\Delta f$ were highly reproducible over weeks: the time elapsed between collection of the magenta/cyan data and the red/blue data in Figs. \ref{fig:PT-damping} and \ref{fig:PT-f} was 37 days. Furthermore, in all but one drum, both $\Delta f$ and $f_m$ were insensitive to changes in pump power. In Fig. \ref{fig:PT-f}, we varied the pump power by over a factor of ten, showing that the pump power is not responsible for decoupling of the TLS. It was also possible to shut off the pulse tube for intervals of about 10 minutes without substantially increasing the warming rate of the nuclear stage. The black (pulse tube on) and orange (pulse tube off) points in the figures show that changing the state of the pulse tube had no effect on the thermal decoupling causing the leveling off of $f_m$. The inset of Fig. \ref{fig:PT-f} shows the time dependence of $f_m$ while the pulse tube was switched on and off. Two measurements of $\Delta f$ and $f_m$ were made each time the pulse tube was shut off.

\begin{figure}
\includegraphics[width=\linewidth]{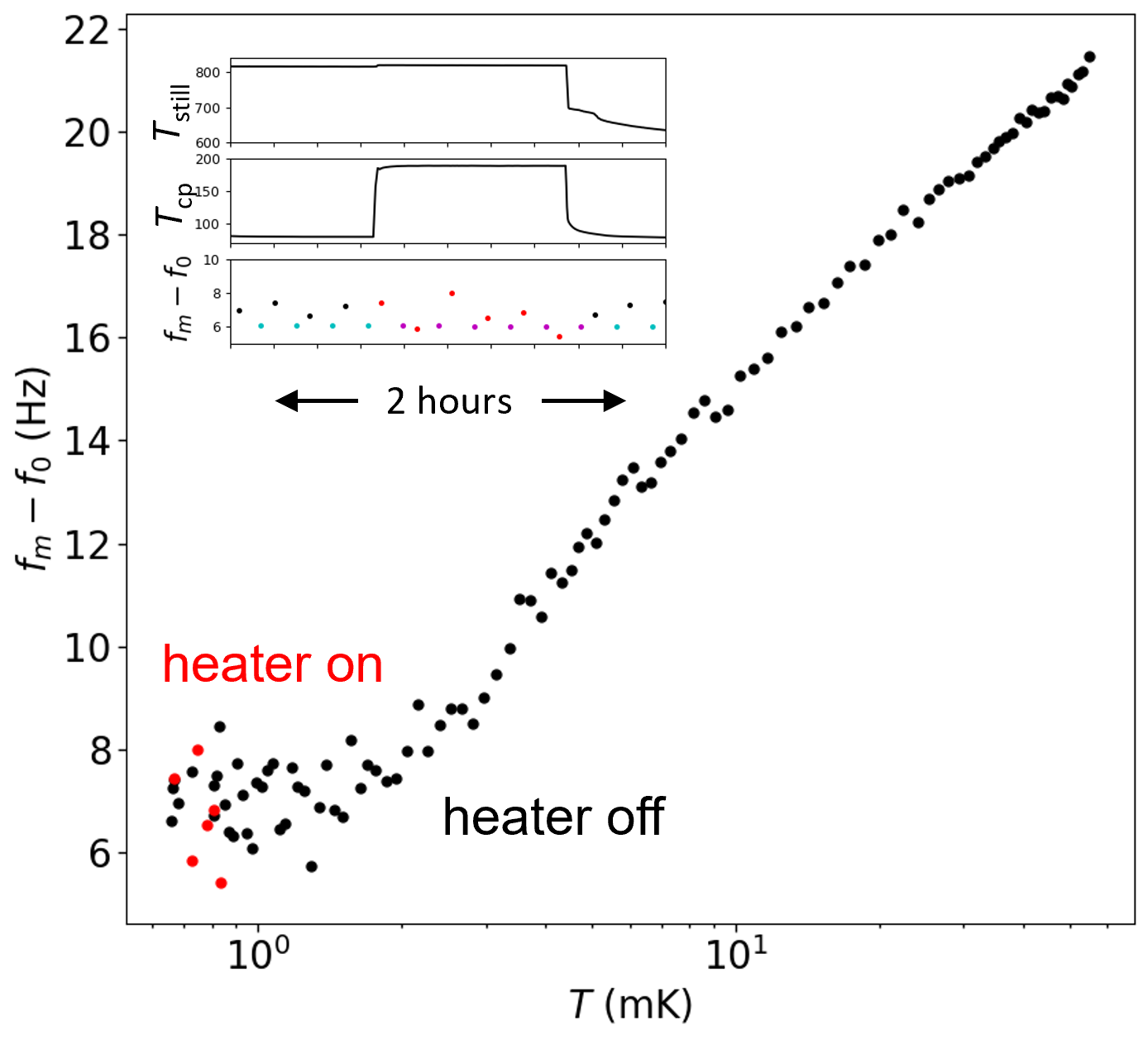}
\caption{\label{fig:heater-f} Main: Frequency shift of a drum vibrating at 1.64 MHz with the cold plate heater turned on (red) or black (off). Inset: Time trace of the still and cold plate temperatures in mK including the interval when the cold plate heater was turned on. The still heat was turned off at the same time as the cold plate heat. The bottom time trace shows the frequency shift in Hz of the 1.64 MHz drum and also the 1.30 MHz drum discussed in Figs. \ref{fig:PT-damping} and \ref{fig:PT-f} (cyan and magenta points).}.
\end{figure}

We also considered the possibility that the upper stages of the cryostat could heat the samples by thermal radiation, via the microwave lines or through free space. In the case of qubits, it is essential to thermalize the modes of waveguides as close as possible to the sample temperature so as to limit decoherence \cite{scigliuzzo2020primary}. Since the coldest attenuator on our microwave pump line was located on the cold plate, it formed a warm bath with a temperature normally near 90 mK when the sample was at base temperature near 1 mK. However, Fig. \ref{fig:heater-f} shows that heating the cold plate to almost 200 mK had no effect on $f_m$ of two drums, one with relatively high $T_{\mathrm{eff}}=6$ mK and the other with low $T_{\mathrm{eff}}=0.7$ mK. The still heater was turned off at the same time as the cold plate heater, causing the still plate to cool, but this also had no effect on $f_m$. Furthermore, turning off the pulse tube caused the 4 K plate to warm from 3.5 K to 7 K before it was turned back on, as well as moderate heating of the 50 K plate, but this also caused no change in $f_m$ (Fig. \ref{fig:PT-f}). We conclude that the thermal decoupling of the TLS responsible for the temperature dependence of $f_m$ is less sensitive to radiation via microwave lines than the thermal decoupling of superconducting qubits. In the present work, propagation of radiation through free space was limited by standard reflective radiation shields as well as the absorbing shields discussed in Sec. \ref{sec:expt}. While radiation from warmer parts of the cryostat does not affect the measurements reported here, colder attenuators or IR filters (e.g. Eccosorb) may be useful once the dominant heat load on our devices is identified and eliminated.

\subsection{Thermalization of a mechanical mode}
\label{sec:mode_therm}
\begin{figure}
\includegraphics[width=\linewidth]{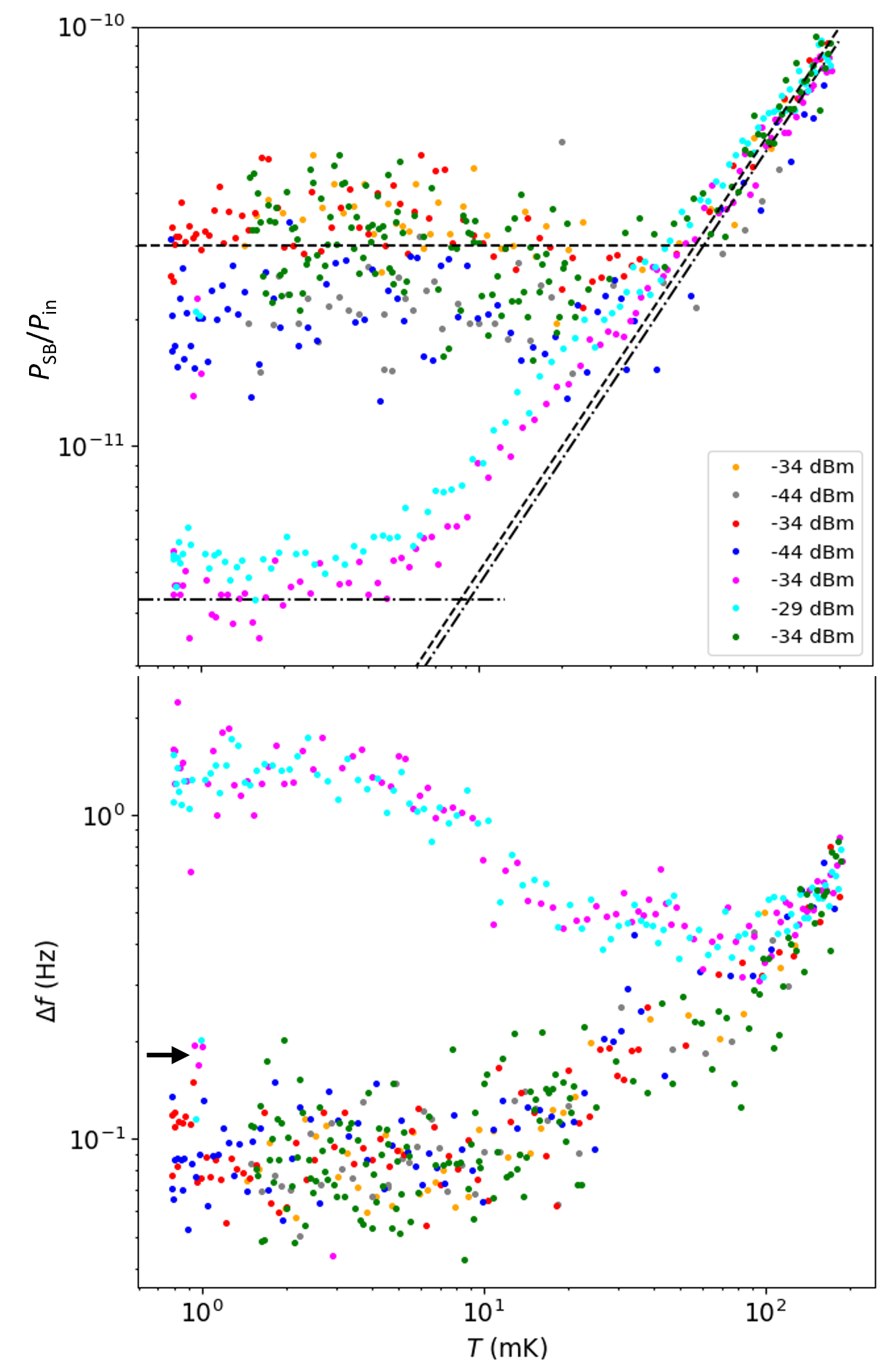}
\caption{\label{fig:area-vs-T} Upper panel: Effective temperature of the 1.30 MHz mode, expressed as $P_{\mathrm{SB}}/P_{\mathrm{in}}$. The pump frequency equaled the cavity frequency and the pump powers are given in the legend. Lower panel: mechanical linewidths corresponding to the measurements in the upper panel. The mechanical mode underwent a temporary transition, and the outlying cyan and magenta points indicated by an arrow are significant (Sec. \ref{sec:mode_therm}).}
\end{figure}

Finally, we discuss the effect of mechanical damping on the phonon occupation of a mechanical mode. For a pump frequency equal to the microwave resonance and in the limit of low pump power $P_{\mathrm{in}}$, the power in the mechanical sideband of the pump is $P_{\mathrm{SB}}=MP_{\mathrm{in}}T_{\mathrm{mode}}$ where $T_{\mathrm{mode}}$ is the effective temperature of the mechanical mode and $M$ depends on the damping of the microwave cavity \cite{zhou2019chip}. The upper panel of Fig. \ref{fig:area-vs-T} shows the temperature dependence of $P_{\mathrm{SB}}/P_{\mathrm{in}}$ for the 1.30 MHz mode already described in Figs. \ref{fig:PT-damping} and \ref{fig:PT-f}. After obtaining each mechanical data point, we maintained the pump settings and probed the cavity, finding that the lineshape of the microwave cavity was stable and nearly independent of temperature. Therefore, $M$ was nearly constant, and $P_{\mathrm{SB}}/P_{\mathrm{in}}$ is a measure of the effective temperature of the mechanical mode. We also verified that our amplifier gains were stable by measuring the probe transmission at the sideband frequency, allowing us to accurately determine $P_{\mathrm{SB}}/P_{\mathrm{in}}$. As expected, $P_{\mathrm{SB}}/P_{\mathrm{in}}$ has a nearly linear dependence on cryostat temperature at relatively high temperatures, demonstrating good equilibration of the mechanical mode to the cryostat. The slanted dashed and dashed-dot lines in Fig. \ref{fig:area-vs-T} are linear fits to the respective green and magenta data above 100 mK.

The measurements below 100 mK in the lower panel of Fig. \ref{fig:area-vs-T} reveal an exception to the excellent reproducibility of $\Delta f$ discussed above. The cyan and magenta data points were obtained by alternating between -34 and -29 dBm pump power as the cryostat warmed from base temperature, and the measured damping was much higher than for all of the other temperature sweeps of this sample. During the temperature sweep that yielded anomalous damping, in a narrow range of temperatures near 1 mK, the damping briefly returned close to its usual low value (see outlying cyan and magenta points), then remained high for the rest of the temperature sweep. The green data were obtained 7 days after the anomalous data, and the other data points were obtained several days before the anomalous data. The upper panel of Fig. \ref{fig:area-vs-T} shows that the effective mode temperature was lower when the mode was in its high damping state. Horizontal lines indicate the average value of the green and magenta data points in the low temperature limit. The intersection of these lines with the linear fits imply saturation temperatures of 9 mK and 60 mK in the high and low damping states, respectively. A jump in $f_m$ occurred while the damping was in the high state, preventing an unambiguous determination of the TLS temperature $T_{\mathrm{eff}}$ using the procedure described above.

The effective mode temperatures of other drums of our sample were sometimes erratic even though $\Delta f$ and $f_m$ were reproducible. Furthermore, the temperatures of some modes were dependent on pulse tube state, in contrast to the absence of an effect of pulse tube state on $\Delta f$ and $f_m$ (Figs. \ref{fig:PT-damping} and \ref{fig:PT-f}). Factors other than the damping are therefore important in determining the relative thermal decoupling of modes of different drums from the cryostat. However, the rare transition of the 1.30 MHz mode described above provided a unique opportunity to study the effect of damping on the thermal decoupling of a single mode.

\section{Conclusion}
We demonstrated our ability to cool the TLS of drums with moderate damping $\approx20$ Hz well below the usual limit of dilution refrigeration. The fact that the damping is temperature dependent below 10 mK in these drums indicates that the internal damping is not hidden by large clamping losses. This provides motivation for theoretical work on the effects of TLS in stressed mechanics in which the dominant thermal phonon wavelength exceeds the thickness or even the diameter of the resonator. Damping due to relaxation of TLS with a broad distribution of energy splittings was not sufficient to explain all the results reported here. We hypothesized that the unusual temperature dependence of the damping at the lowest temperatures is due to a surplus of defects with a particular energy splitting. Future theoretical work may also explain the exceptionally low tunneling strength $C$ in these devices (Sec. \ref{sec:ThermTLS}) by a dependence on stress and aspect ratio (Sec. \ref{sec:theory}) or in terms of an anomalously low density of TLS.

While the variation in the saturation temperature $T_{\mathrm{eff}}$ of $f_m$ of the different drums could be due to differences in intrinsic behavior, we argued that it is more likely due to differences in thermalization of TLS to the cryostat. Concerning the origin of the thermal decoupling, we did not find evidence for variation in the heat load on the drums due to heat release from TLS, so we considered the thermal resistance between the TLS and the substrate. The low clamping loss and weak coupling between the modes of a drum imply that the dominant thermal path between the TLS responsible for shifts in $f_m$ and the substrate could be due to interactions between TLS. The dependence of the population of one of the mechanical modes on its damping state is consistent with this picture: the dissipation in the state of high damping and low phonon occupation was temperature dependent and therefore not dominated by clamping loss.

The present work exposes a potential challenge in the optimization of mechanical coherence, which requires minimization of the product of damping and equilibrium phonon occupation of mechanical modes. In the devices studied in the present work, the heat load on the drums is apparently high enough so that low damping of the fundamental mode causes it to thermally decouple from the cryostat. However, thermal decoupling is not guaranteed to worsen as mechanical damping decreases: eliminating TLS simultaneously decreases damping as well as the heat load on the mechanics due to TLS heat release. Future work will include measurements of other types of low damping drums that may experience a lower heat load, leading to good thermalization to the cryostat even at sub-mK temperatures.

\section{Acknowledgements}
We acknowledge support from the ERC StG grant UNIGLASS (Grant No. 714692). This work was supported by funding from the Swiss National Science Foundation under grant agreement No. 231403 (CoolMe) and by the Laboratoire d'excellence LANEF (ANR-10-LABX-51-01). The research leading to these results received funding from the European Union’s Horizon 2020 Research and Innovation Program under Grant Agreement No. 824109. We thank Sebastien Pairis for assistance with scanning electron microscopy.

\section{Data Availability}
The data that support the findings of this article are openly available\cite{data}.

\end{document}